ADVANCING EARTH AND SPACE SCIENCES

JGR | Machine Learning and Computation

**RESEARCH ARTICLE**
10.1029/2024JH000366

# Rapid Automated Mapping of Clouds on Titan With Instance Segmentation

Zachary Yahn[1,2] , Douglas M. Trent[3], Ethan Duncan[4], Benoît Seignovert[5] , John Santerre[2], and Conor A. Nixon[2] 

[1]School of Computer Science, Georgia Institute of Technology, Atlanta, GA, USA, [2]Planetary Systems Laboratory, NASA Goddard Space Flight Center, Greenbelt, MD, USA, [3]Office of the Chief Information Officer: Informatio, Data & Analytics Services, NASA Langley Research Center, Hampton, VA, USA, [4]School of Information, University of California, Berkeley, Berkeley, CA, USA, [5]Observatoire des Sciences de l'Univers Nantes Atlantique, Osuna, UAR-3281, Nantes, France

**Abstract** Despite widespread adoption of deep learning models to address a variety of computer vision tasks, planetary science has yet to see extensive utilization of such tools to address its unique problems. On Titan, the largest moon of Saturn, tracking seasonal trends and weather patterns of clouds provides crucial insights into one of the most complex climates in the Solar System, yet much of the available image data are still analyzed in a conventional way. In this work, we apply a `Mask R-CNN` trained via transfer learning to perform instance segmentation of clouds in Titan images acquired by the Cassini spacecraft—a previously unexplored approach to a "big data" problem in planetary science. We demonstrate that an automated technique can provide quantitative measures for clouds, such as areas and centroids, that may otherwise be prohibitively time-intensive to produce by human mapping. Furthermore, despite Titan-specific challenges, our approach yields accuracy comparable to contemporary cloud identification studies on Earth and other worlds. We compare the efficiencies of human-driven versus algorithmic approaches, showing that transfer learning provides speed-ups that may open new horizons for data investigation for Titan. Moreover, we suggest that such approaches have broad potential for application to similar problems in planetary science where they are currently under-utilized. Future planned missions to the planets and remote sensing initiatives for the Earth promise to provide a deluge of image data in the coming years that will benefit strongly from leveraging machine learning approaches to perform the analysis.

**Plain Language Summary** Although deep learning models continue to be popular tools for addressing a variety of computer vision tasks, planetary science has yet to see their extensive utilization. One such application is cloud tracking in Titan's atmosphere, which has dynamic and complex weather phenomena. In this work, we train a deep computer vision model to identify these clouds from images of Titan's surface. We assess the accuracy of this model, and we also show that our approach is more efficient than human labelers. In addition to cloud detection, we also develop codes for automatically extracting data such as cloud areas and latitudes from our model's predictions. We also propose that similar methods have broad applicability to other planetary science problems, especially given several upcoming missions that promise an influx of high-quality image data.

## 1. Introduction

### 1.1. Titan's Meteorology

Titan, Saturn's largest moon, has an atmosphere denser than that of any other moon in the Solar System. This atmosphere is composed mostly of nitrogen ($N_2$, 95%–98% by altitude), with a significant minority of methane ($CH_4$, 5%–2%) (Niemann et al., 2010). Methane on Titan takes on a similar role to water in the Earth's meteorology (Hörst, 2017). It is evaporated from the surface, and humid near-surface air parcels rise until the atmosphere is cold enough (15–30 km) for droplets to condense (Mitchell & Lora, 2016). In the stratosphere, ice clouds are seen at different altitudes and seasons which may be composed of other organic molecules besides methane: HCN, $C_2H_6$, $C_4N_2$, and $C_6H_6$ clouds have all been proposed based on condensation altitude (Anderson et al., 2016; de Kok et al., 2014; Griffith, 2006; Vinatier et al., 2018), however these identifications remain uncertain.

Achieving a definitive understanding of these clouds may ultimately require in situ measurements, but while we await future missions to achieve these goals, more sophisticated modeling constrained by the available data from the Cassini-Huygens provides an avenue to make progress. To this end, a first step is making a complete mapping of cloud locations, sizes and shapes across the mission from the visible camera data.

Clouds on Titan are known to possess a variety of shapes: at times they may appear patchy, streaky or wispy, similar to clouds on the Earth. A few large storm fronts have also been seen (Turtle et al., 2011, 2018). Clouds migrate from hemisphere to hemisphere, appearing mostly in the summer hemisphere. This trend is occasionally disrupted, such as a south polar vortex seen at the start of southern winter (Turtle et al., 2018).

Understanding the seasonal weather patterns of Titan's clouds may also lend greater understanding into the atmospheric processes of other Solar System worlds including our own (Lora et al., 2015; Mitchell et al., 2011). However, obtaining a high-quality data set of cloud locations, morphologies and other information comes with a host of challenges.

First, capturing high-quality images of clouds on Titan is not straightforward. The moon's atmosphere possesses a haze that is opaque to optical observations, excepting only specific ranges of the infrared spectrum (Turtle et al., 2011). Cassini's ISS-NAC (Imaging Science Subsystem Narrow Angle Camera), one of two cameras used to collect near-infrared images on the spacecraft, included several infrared filters capable of piercing the methane haze and enabling capture of cloud images (Porco et al., 2004). While the cloud detection problem on Earth benefits from verification from multiple sources and spectral ranges, as utilized by Letu et al. (2022), Cassini's ISS was only capable of using a single infrared filter at a time. Because the Saturnian year, and thus Titan's, is 29.4 Earth-years long, Titan's seasons last significantly longer than those on Earth (Mouelic et al., 2018; Rodriguez et al., 2009, 2011). For this reason, cloud development must ideally be observed over several Earth-decades to acquire a complete annual cycle. Consequently, the Cassini data set built up over 127 Titan targeted flybys from 2004 to 2017 is the richest and most extensive collection of Titan images acquired to date. While ground-based telescopes provide more temporal/spectral coverage, Cassini was capable of resolving individual clouds. Figure 1 shows examples of cloud-containing and cloud-free images acquired by the Cassini ISS.

Planetary scientists have labored to manually comb through the Cassini image data to establish basic trends and features of the clouds (Turtle et al., 2018), but there is still much to uncover. Researchers are interested in tracking cloud areas, shapes, locations and local times over multi-year periods. While this kind of specific information is fairly onerous for manual tabulation over the entire mission, and thus its collection has never been attempted, a rapid, automated method of deriving cloud parameters would provide a critical resource that may be used to improve the accuracy of atmospheric models of Titan (Brown et al., 2010).

## 1.2. Machine Learning Approaches Already in Use for Planetary Atmospheres

Recent advances in remote sensing research of the Earth's atmosphere can provide insights into how automated cloud detection might be approached on other worlds. The vast majority of cloud identification research focuses on terrestrial applications due to the immediate relevance to human society and multitudinous high-quality data sources. Before the popularization of deep learning, classic approaches such as handcrafted feature representations saw success. While capable in some cases, these models struggled to generalize to the diversity of Earth weather, a substantial drawback given the wide variety of cloud formations possible on Earth (L. Li et al., 2021; Mahajan & Fataniya, 2020). Deep computer vision models, including fully convolutional networks, have proven to be more effective in side-by-side comparisons (LeGoff et al., 2017). Several specialized approaches have been implemented, including multi-scale convolutional feature models that incorporate multiple layers of the same image from different sensors (e.g., RGB and infrared) (Z. Li et al., 2019), ensemble approaches with independent feature extraction and boundary refinement modules (Chen et al., 2021), and many others. Semantic segmentation architectures are also popular, especially the `U-Net` (Francis et al., 2019) and `ResNet` (Mommer, 2020) architectures. Semantic segmentation involves producing a mask that labels the individual pixels of an image. These techniques have demonstrated per-pixel accuracies as high as 0.9 for terrestrial cloud applications (Francis et al., 2019; LeGoff et al., 2017; L. Li et al., 2021; Mommer, 2020). Some deep learning approaches have even become advanced enough to target specific cloud formations. Notably, Liles et al. (2020) developed a U-Net model for detecting Above Anvil Cirrus Plume (AACP) formations, a reliable predictor of imminent severe weather. Other approaches demonstrate the feasibility of cross-scene hyperspectral classification (Zhang, Li, Tao, et al., 2021; Zhang, Li, Zhang, et al., 2021; Zhang et al., 2022). These significant advances in cloud detection are possible because of the vast quantities of image data available through several satellites orbiting the Earth, enabling the full capabilities of deep learning.

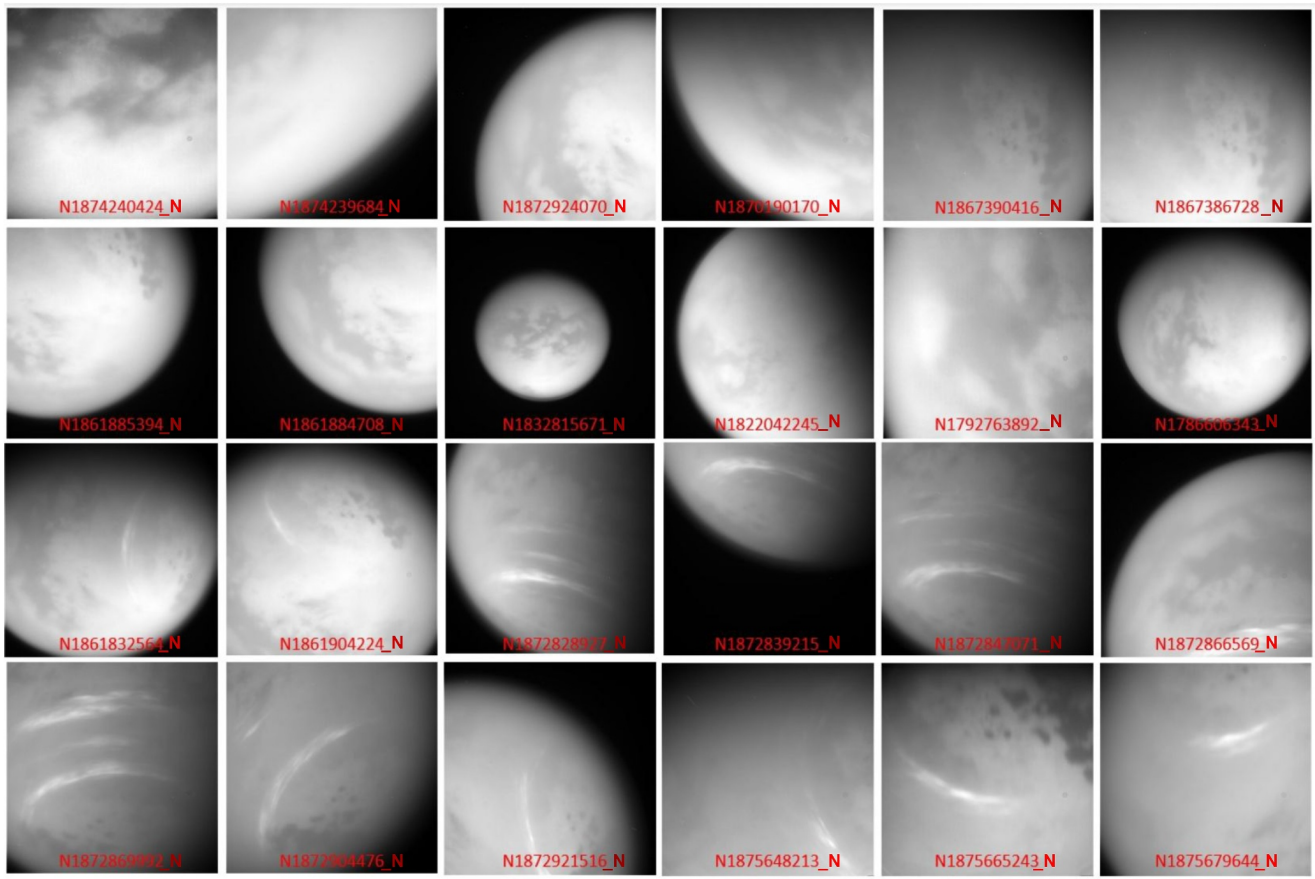

**Figure 1.** Examples of images in the Cassini ISS data set taken with the Narrow Angle Camera (NAC) CLR-CB3 filters: upper two rows—no clouds. Lower two rows—with clouds. Note varying viewing angles, surface sections, and distances for each capture. Image names as listed in NASA planetary data system (PDS) provided (JPL et al., 2017).

Deep learning has revolutionized computer vision by using very large networks with dozens of layers to model rich relationships between individual features (LeCun et al., 2015). At the present day, some of the most widely known and utilized architectures are adaptations of Convolutional Neural Networks (CNNs), with particular application to image classification and object detection tasks (Krizhevsky et al., 2012). However, designing and coding a new deep learning model is difficult, time consuming, data-intensive, and therefore often beyond the scope of a typical project's time and resource budgets. Popular architectures often involve tens or hundreds of millions of parameters, resulting in massive computational costs for even the simplest training.

By taking advantage of similarities between domains, transfer learning addresses problems with limited data. Transfer learning uses abstract features such as curves or edges learned from one foundational domain and adapts them to another. Because many of the best pre-trained models target a broad array of classes via data sets like `ImageNet`, their early layers are applicable to a wide range of computer vision problems. Transfer learning takes advantage of this to preserve early learned features by "freezing" those layers and only training the last few layers of a network on a novel data set. This is advantageous for small budget projects, as it allows for the utilization of large, well-trained models with pre-trained weights to be tuned on a minimal amount of new data with few additional training epochs. In summary, transfer learning reduces the dependency on large training data sets and minimizes training time (Zhuang et al., 2020). Transfer learning has been shown to be especially efficacious in computer vision problems (X. Li et al., 2020). Example use cases include forestry (Kentsch et al., 2020), pavement fault detection (Gopalakrishnana et al., 2017), and facial recognition (Cao et al., 2013). For these reasons, we explored transfer learning as an efficient way to train models for Titan cloud identification.

### 1.3. Application to Titan

Image-based Titan cloud identification is especially challenging due to the high variability of cloud shapes, sizes, opacity, latitudes, and the lack of contrast with the surface of Titan. These difficulties are illustrated in the example images in Figure 1. Other planets present similar difficulties and opportunities. With the exception of Mercury, every planet in the Solar System has a dynamic atmosphere that results in cloud formations. Each has a distinct surface colorization, texture, and exhibits other conditions that make it unique. Designing individual machine learning models for each world, many of which have very limited data available, is infeasible. While some previous publications mention computerized cloud detection (e.g., Turtle et al., 2018), the only dedicated example in the literature involves using a Bayesian Source Separation algorithm paired with Markov Chain and Monte Carlo simulation methods (Mouelic et al., 2018; Rodriguez et al., 2010). Therefore, there is considerable scope for further application of machine learning to planetary cloud detection and other science areas, given the surge in development of new deep learning techniques in the last decade.

### 1.4. Contribution and Significance

Although machine learning has been applied in recent years to some of NASA's key focus areas—for example, spectroscopy (Wilkins et al., 2020), rover vision systems (Dundar et al., 2019) and crater detection (Di et al., 2014)—its adoption in the planetary science community has been slower than in other areas of space science, perhaps due to historically smaller data sets. In a white paper for the *NRC Planetary Science and Astrobiology Decadal Survey*, 51 researchers signed a statement recognizing a need for machine learning to be more widely deployed in planetary science projects (Azari et al., 2020). They note that the proportion of planetary science papers that include machine learning is less than half of that in the other three NASA science areas: heliophysics, astrophysics, and Earth science.

In this paper, we exploit and demonstrate how a deep learning computer vision approach can deliver accurate and efficient models for Titan cloud identification based on a subset of the available data. To our knowledge, our team is the first to implement a deep learning computer vision model for the purpose of Titan cloud feature recognition. We first implement an instance segmentation model to identify whether or not a Cassini image of Titan contains one or more cloud formations, and which image pixels are contained within the cloud. This discrimination is essential for determining the spatial distribution of clouds over an extended time period, which is a topic of considerable interest for research into Titan meteorology (Rodriguez et al., 2010; Turtle et al., 2018). We then use the masks generated by our instance segmentation model to calculate the areas, centroids, and aspect ratios of Titan clouds, providing valuable metrics that were previously too labor-intensive to be calculated by hand. Furthermore, we show that this approach is faster than manual tabulation, and consistent with contemporary applications to other contexts. In addition to developing a model that effectively identifies clouds on Titan from historic Cassini data, we envision that future data pipelines may be implemented onboard spacecraft to process real-time data from missions to Titan and other planets. The ability to create higher-level data products on board spacecraft would in turn enable substantial reduction in data downlink requirements.

## 2. Materials and Methods

In this section we present a novel cloud identification model for addressing the challenges outlined in Section 1 by classifying images with a `Mask R-CNN` (Regional Convolutional Neural Network), a popular image classification architecture (He et al., 2017). We provide an overview of this strategy, our data set, and our transfer learning implementation.

### 2.1. Data Set

Cassini ISS image data are publicly available on the Jet Propulsion Laboratory's Cartography and Imaging Science Node of the Planetary Data System (JPL et al., 2017). The initial scope of this work encompassed all images of Titan captured by the ISS for the full duration of Cassini's Titan viewing window (June 2004–October 2017). However, the vast majority of these images are not suitable forour work because Titan's methane haze obfuscates any cloud structure. The ISS provides multiple filters, up to two of which can be used at a time for each of the Narrow Angle Camera (NAC) and Wide Angle Camera (WAC). The CL1-CB3 filter, which is centered at 938 nm (Porco et al., 2004) is ideal for capturing images of Titan's clouds. At this wavelength, clouds are clearly discernible as high-albedo shapes against the less reflective surface of Titan, though their visibility still varies widely depending on factors such as proximity to the planetary limb because this wavelength pierces Titan's methane haze. For this reason, we only select images captured using the CL1-CB3 filter.

Prior work has also shown that clouds were not dispersed evenly throughout the mission time window, but rather appeared in high densities during specific months (Rodriguez et al., 2011; Turtle et al., 2011). To further narrow the scope of this work, we selected a window of September 2016–September 2017 in order to capture one of these high-density periods. From this window we selected 1,053 total images. These images are intended to be a representative sample of the wide variety of cloud types and image conditions in the Cassini data. We further separated these 1,053 images into training and testing sets at random without replacement: 642 for training, and 411 for testing. The training set is further randomly split into training and validation sets, with 428 images in the training set and 214 in the validation set. The validation set was used to assess performance during training while saving the testing set to evaluate generalizability. All ground truth labels were produced manually with LabelMe (Wada, 2024). Labels were bounding polygons where every pixel enclosed in the polynomial is considered part of the cloud. Multiple unique instances of clouds each had distinct labels within a single image. Each image is either $1024 \times 1024$ or $512 \times 512$ pixels. We use 8-bit uncalibrated images to show that the model can identify clouds from raw data, which may be important in situations such as live processing onboard a spacecraft. However, it should be noted that researchers who work with images of Titan use the 12-bit calibrated versions, which account for possible trends induced by measurement error. All images were resized to $512 \times 512$ for consistency. We provide our full data set online as described in Data Availability Statement.

Several factors make instance segmentation on the Cassini data set a challenging task. Being in orbit around Saturn, the distance between Cassini and Titan varied significantly throughout the course of the mission. While Earth-based satellites typically orbit at a constant planeto-centric distance, giving images a consistent pixel scale, Cassini followed an orbit around Saturn that brought it from less than 5,000 km to greater than 5,000,000 km from Titan during the mission. Thus, its images were captured from diverse viewing angles with disparate lighting conditions and resolutions. We summarize the challenges with the Cassini image data set as follows:

1. Cloud shape and size variability, ranging from narrow streaks to pseudo-elliptical masses;
2. Serial, monochromatic image data captured from the CL1-CB3 filter centered at 938 nm wavelength.
3. Lack of cloud-surface contrast, especially toward high emission angles (close to the planetary limb);
4. Cassini's varying distance to Titan, creating a non-uniform pixel scale between different images.
5. During any given flyby only a fraction of Titan is visible.
6. Sparse and irregular temporal coverage.

The narrow subset of images we select for this work is an attempt to make the problem tractable without losing the challenging attributes of the whole data set. We use a curated sample to show that instance segmentation is a viable approach to this sort of problem, and also to highlight the applicability of transfer learning to tasks with small data sets.

## 2.2. Model Architecture and Training

Our model is a Mask Region-based Convolutional Neural Network (`Mask R-CNN`) (He et al., 2017) with a Residual Network 50 (`ResNet50`) (He et al., 2015) backbone. Although more advanced instance segmentation models exist, especially those based on transformers, we choose `Mask R-CNN` because it is lightweight and easily accessible through PyTorch (PyTorch, 2017). This makes it ideal for other researchers who may not have experience with implementing deep learning models. The `Mask R-CNN` first uses the `ResNet50` backbone to extract descriptive features from the image using successive groupings of convolutional layers with residual connections. Then, the `Mask R-CNN`'s Region Proposal Network (RPN) identifies significant regions of potentially relevant shapes in the latent feature space. Within each region identified as a cloud, the `Mask R-CNN` prediction head then maps these features back to image pixels and assigns each pixel a label, creating a segmentation mask for every cloud instance. This architecture is shown in Figure 2.

The model weights were pre-trained on the Common Objects in Context (COCO) data set (Lin et al., 2014) to learn general feature detection for a wide variety of objects. COCO contains over 14 million images relating to tens of thousands of classes. Specifically, we use the pre-trained `Mask R-CNN` available through PyTorch (PyTorch, 2017). Per a transfer learning approach, we then tuned this pre-trained version to our specific use case, applying our own training using our pre-selected and labeled subset of Cassini images. During this training all weights in the model were frozen except for the prediction head. We trained with a binary cross-entropy loss function, a batch size of 16, and a learning rate of 0.00001. The Adam optimizer (Kingma & Ba, 2014) was used to improve performance during training. Training converged after 20 epochs, as visible in Figure 3.

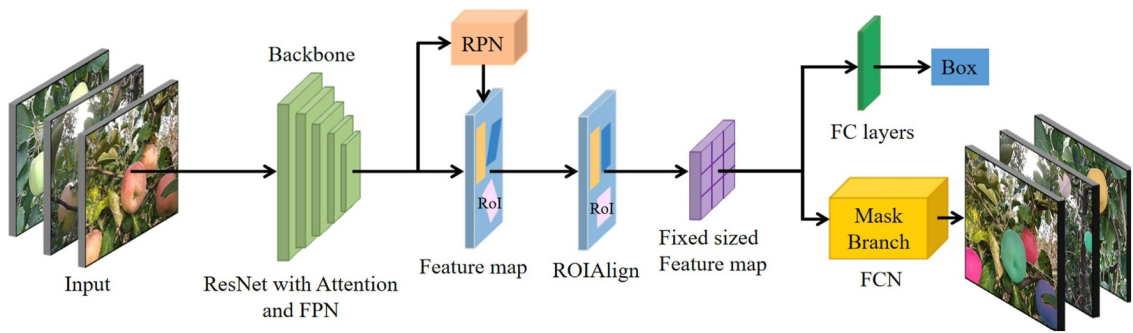

**Figure 2.** Architecture of `Mask R-CNN`, a widely used and well documented model (Wang & He, 2022). We employ our `Mask R-CNN` tuned via transfer learning for instance segmentation on our Cassini clouds data set. Note that this visualization is a slight variation of the original model.

### 2.3. Metric Calculation

Given a segmented mask output for each input image, our goal was to build a framework (Yahn et al., 2024b) for calculating metrics such as cloud area, centroid coordinates, and aspect ratio, with the potential for additional metrics in the future based on input from other researchers. The acquisition geometry (data for each image that include Cassini's position and angle relative to Titan) was retrieved using the same approach as in Seignovert et al. (2021) using the NAIF toolkit and Cassini SPICE kernels (Acton, 1996; Acton et al., 2017). Latitude and east longitude are stored for each pixel in backplanes. These data are specific to each image, meaning they account for the varying angle and distance of Cassini with respect to Titan throughout the mission. Although for our purposes these data were only collected for the images in the test set, they are available for all Cassini images.

Per-pixel latitude and east longitude backplanes enable the calculation of per-pixel areas for the predicted mask. Per-pixel areas are summed to compute the total area of the cloud. The cloud's centroid is calculated via a weighted center of mass over every pixel in the mask, where each pixel is weighted by its area. Finally, the aspect ratio is calculated by finding the major axis of the cloud that is, the line segment connecting the two most distance points in the mask, and then finding the two furthest points on the minor axis perpendicular to this major axis. The ratio of the lengths of these axes is the aspect ratio. Areas, centroids, and aspect ratios are calculated in terms of latitude and east longitude angles to account for the curvature of Titan. A full tabulation of metric calculation results for every predicted cloud instance is provided as supplementary material (https://zenodo.org/records/14009768).

### 3. Results

In this section we present our results from using the methodology described above to train the model, and then applying the trained model to our testing data set. We show the outcome of calculating the area, centroid, and aspect ratio of every cloud in the testing set.

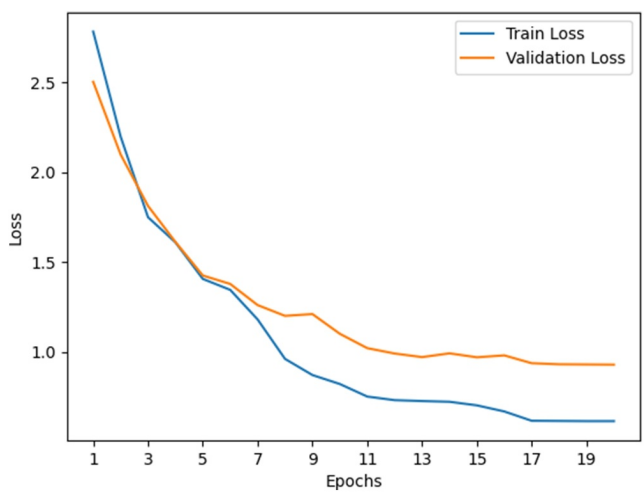

**Figure 3.** Binary cross-entropy (BCE) losses on the training and validation sets during learning. BCE is a dimensionless quantity that measures how well the model classifies each pixel in an image, calculated over every image in the respective set. Lower values mean fewer incorrect per-pixel predictions, treating false positives and false negatives equally. The smooth decrease of both curves shows that our model converges during training, and the small gap between validation loss and training loss indicates minimal overfitting.

### 3.1. Segmentation

We first evaluate model performance on the basis of how well it identified individual instances of clouds, not on a per-pixel level. For example, if an image contains three instances of clouds, we measured whether or not the model correctly identified the presence of all three. To quantify this, we make use of accuracy, precision, and recall. Accuracy reflects how many total guesses, positive and negative, are correct. Precision denotes how often a positive guess is correct. Recall indicates, given that a cloud exists in an image, how likely the model is to detect and correctly classify it. These are all calculated as follows (TP is true positive, TN is true negative, FP is false positive, FN is false negative):

$$\text{Accuracy} = \frac{TP + TN}{Total} \qquad (1)$$

$$\text{Precision} = \frac{TP}{TP + FP} \qquad (2)$$

$$\text{Recall} = \frac{TP}{TP + FN} \qquad (3)$$

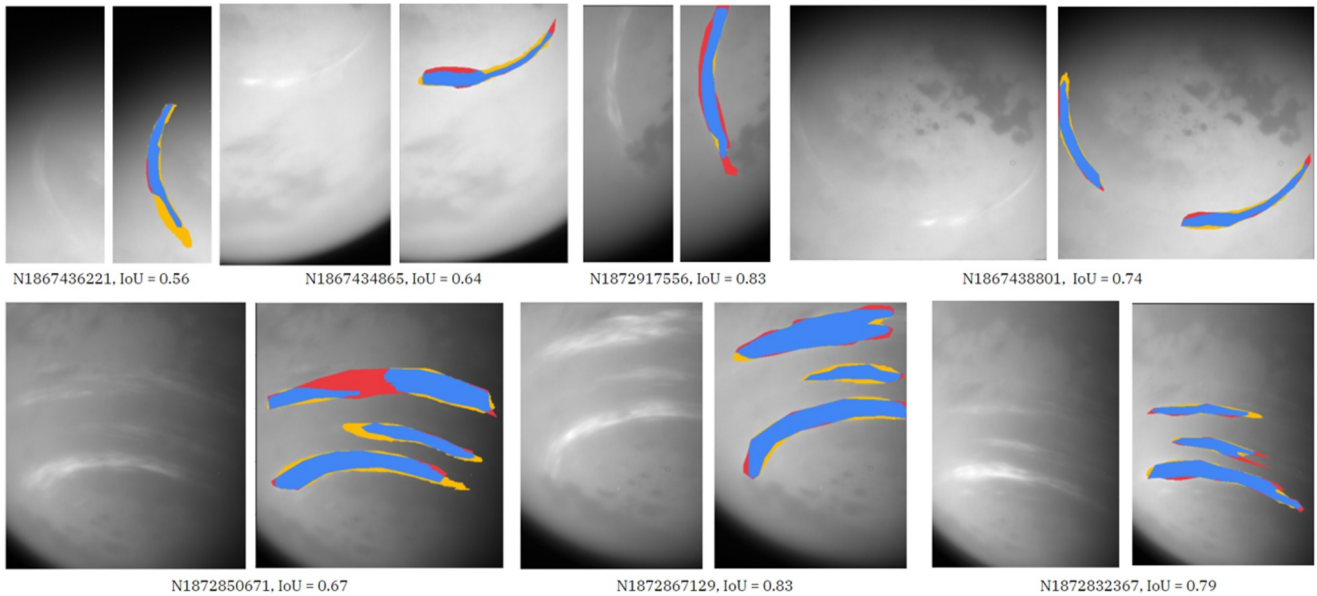

**Figure 4.** Instances from the test set where the model struggled to find precise edges of the clouds, demonstrating a potential source of error. Blue pixels are true positives, red pixels are false negatives, and yellow pixels are false positives. True negatives are left as background because they comprise the vast majority of all pixels.

On the 411 images in the test set, the model achieved an accuracy of 0.96, a precision of 0.80, and a recall of 0.88.

We then evaluate how well the model classified pixels in a given image, contributing to the overall efficacy of the segmented mask output. We make use of precision and recall, and of the intersection over union (IoU) score, which is calculated as follows:

$$\text{IoU} = \frac{TrueMask \cap PredictedMask}{TrueMask \cup PredictedMask} \qquad (4)$$

Thus the minimum IoU score is 0, when the output mask shares no overlap with the true mask (or the model guesses that there is no cloud in the image when in fact there is), and the maximum is a score of 1, when the output mask and true mask are completely aligned. This IoU score was calculated for each image where the model predicted the presence of at least one cloud. In many cases, the model correctly identified the general location of the cloud formations in an image, but struggled to precisely outline the edge of the cloud structure. Even for a human, identifying where these edges stop and the atmospheric haze begins is challenging. This struggle with edge detection likely contributed to a lower IoU score. Examples of these cases are shown in Figure 4.

Note that the mask inferred by the `Mask R-CNN` includes a confidence value for each pixel ranging from 0 to 1. Establishing a clear boundary for the predicted mask requires choosing a threshold above which a pixel is considered part of the cloud. Figure 5 shows the average precision, recall, and IoU for the cloud masks at various thresholds for clouds in the validation set. Note that these precision and recall values are different from the above results, which were based on identifying the presence of clouds. Rather, these scores report the efficacy of the mask at classifying individual pixels. One can see that the IoU is greatest at a threshold of 0.5, which is also where precision and recall are most balanced. Thus, we choose this threshold for our metric extraction.

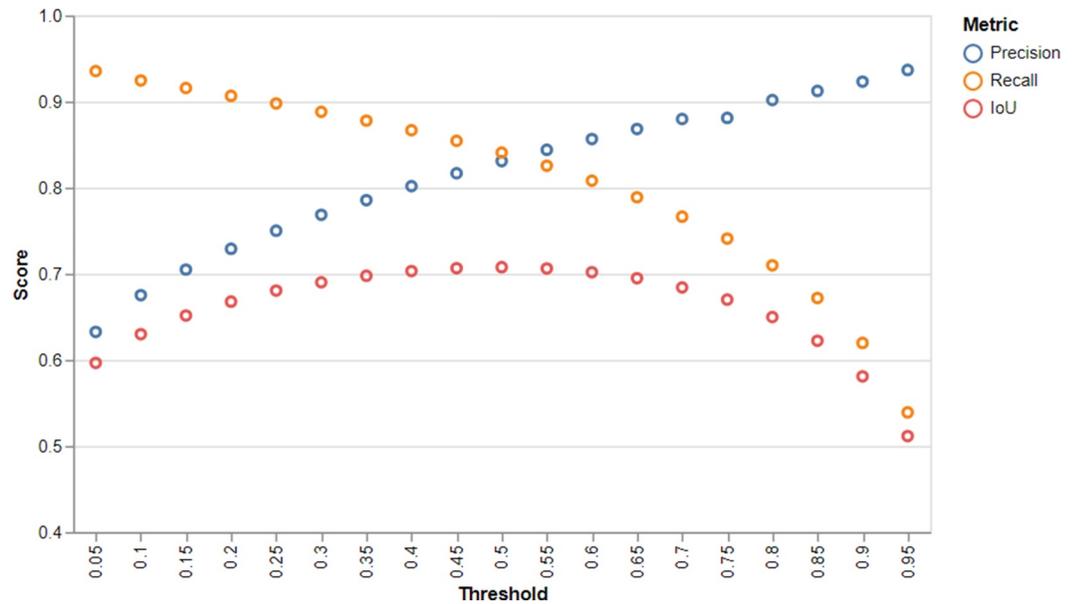

**Figure 5.** Average IoU, Precision, and Recall calculated over the test set at various pixel confidence thresholds. Lower thresholds intuitively encourage more liberal predictions, increasing recall but harming precision. Conversely, higher thresholds beget conservative predictions, increasing precision at the expense of recall. We select a threshold of 0.5 for our metric extraction because it strikes a balance between this precision and recall tradeoff, and because it maximizes IoU.

### 3.2. Cloud Metrics and Trends

We also generated a series of graphs to illustrate how our automated inference of cloud pixel masks and attributes such as area can uncover trends. For example, Figure 6 shows the areas, latitudes, and aspect ratios of clouds during our selected time window (September 2016–September 2017). During 2017, aspect ratios tended to stay constant, with the majority of clouds being long and thin. Likewise, centroid latitudes remain relatively constant around ∼50° throughout that time period, as seen in previous studies (Turtle et al., 2018). Note that the clouds cluster around a few epochs because these are the dates of Cassini's Titan flybys.

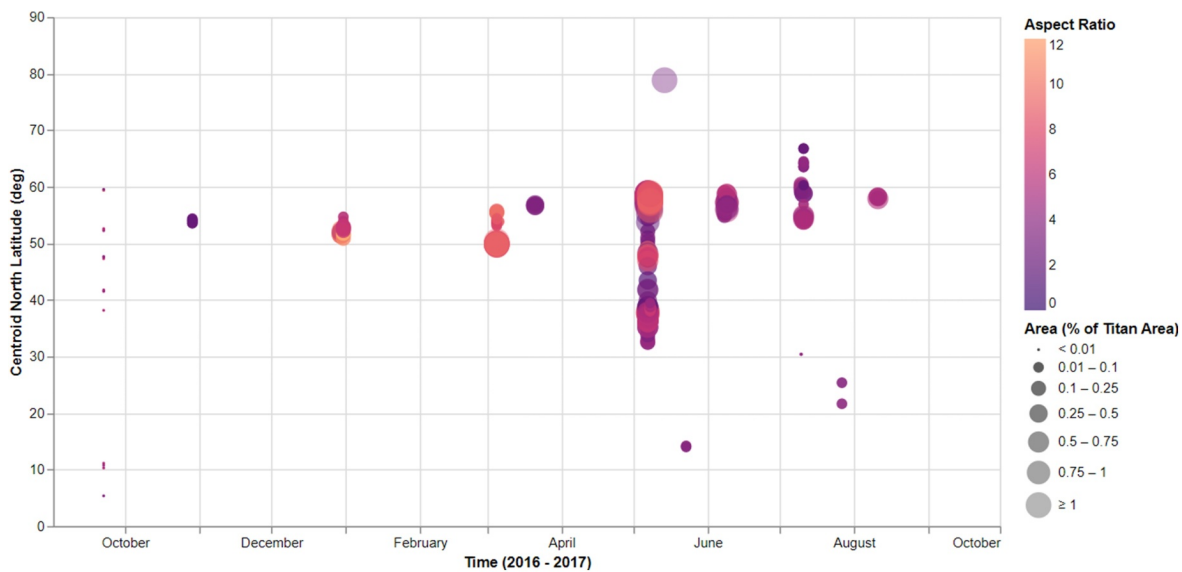

**Figure 6.** Areas, north latitudes, and aspect ratios of clouds during the 2016–2017 time window we selected for our data set. Cloud observations are most prevalent in May 2017, displaying varying aspect ratios and areas. Note that vertical clustering is the result of discrete Cassini flyby windows.

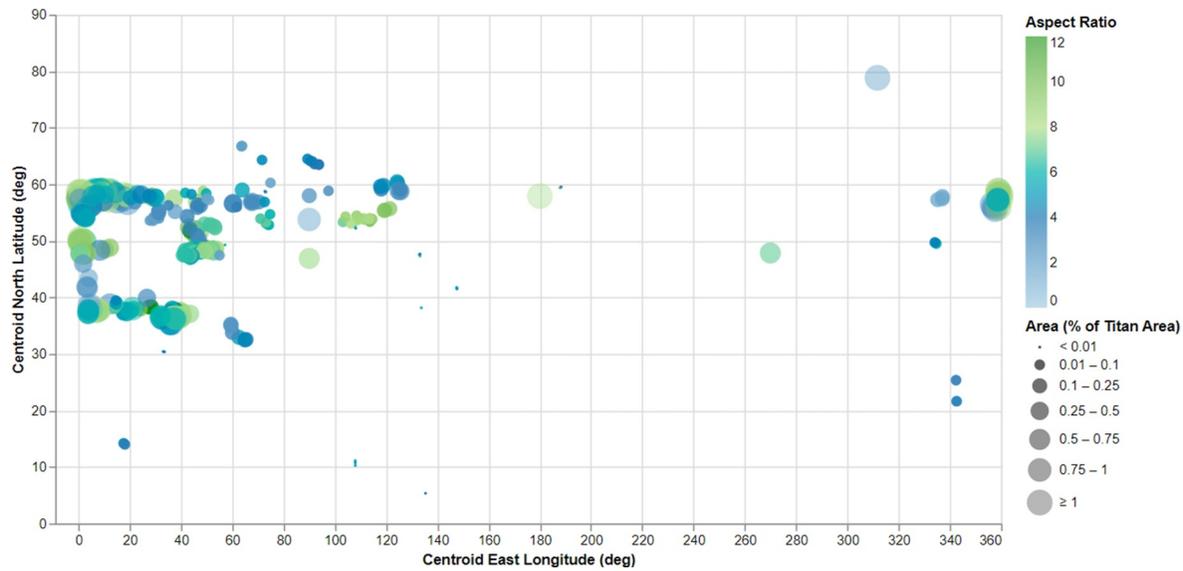

**Figure 7.** Distribution of areas, north latitudes, east longitudes and aspect ratios of clouds during the 2016–2017 time window. Clouds in this time frame tend to cluster between 30 and 60° north and 340 to 130° east, with a few outliers. They display a wide variety of aspect ratios and areas.

This approach can also visualize summary trends over the course of a given time window. For example, Figure 7 shows where clouds tended to be located, as well as their sizes and aspect ratios. One can see that during our selected time window clouds tended to be around ∼50° latitude and between ∼340° and ∼120° east longitude. Both Figures 6 and 7 are consistent with human-mapped clouds, most notably in Turtle et al. (2018).

Note that these figures are intended as examples; complete analysis of the trends would require applying our approach to a larger sample size of the Cassini data. Our goal is to demonstrate an efficient and accurate tool for producing such metrics, so that other studies could extend this work in the future.

### 3.3. Speed Comparison of Machine Learning Versus Manual Approach

To understand the speed at which our methods can compute information of interest, we recorded how much time was required on average to perform inference on an image and then extract said metrics. Averaged over all 411 images of our testing set, computing the segmented mask for a 512 × 512 image required 1–2 s on a single NVIDIA T4 GPU. Based on the time required to label the training data, we estimated that drawing a label of similar quality would take at least 30 s for a trained human, potentially longer for more complicated cloud geometries. For images that contain an instance of clouds, computing area, centroid, and aspect ratio required an average of around 150 ms with an Intel i7 CPU. To our knowledge, nobody has attempted to manually compute these metrics, so it is difficult to compare this with human rates.

At these speeds, generating labels for all 14,335 images provided by the CL1-CB3 filter would take approximately 8 hr. Generating metrics for these images would take an additional 35 min in the worst case. Although it is difficult to compare this to human performance, the obvious advantages of running this processing without supervision improves upon human efforts. Furthermore, there is potential for significant speedups by applying multi-threading, parallelization, and faster hardware. It is also worth noting that this comparison may change if the model were deployed on-board a spacecraft, although GPUs and TPUs for space applications are currently in development.

## 4. Discussion

Our results indicate that automated cloud identification of images of Titan from the Cassini mission is tractable, and that instance segmentation is an effective approach. We have shown that during training, the model converges

**Table 1**
*Comparison With Earth-Based Cloud Detection Studies for Reference*

| Study | Method | Accuracy | Precision | Recall | IoU |
|---|---|---|---|---|---|
| LeGoff et al. (2017) | CNN | 0.86 | 0.75 | 0.81 | N/A |
| L. Li et al. (2021) | U-Net | 0.96 | N/A | N//A | 0.90 |
| Francis et al. (2019) | U-Net | 0.92 | N/A | N/A | N/A |
| Lopez-Puigdollers et al. (2021) | FCNN | 0.94 | N/A | N/A | N/A |
| Fabel et al. (2022) | U-Net | 0.86 | 0.78 | 0.85 | 0.80 |
| Gonzales and Sakla (2019) | U-Net | 0.96 | 0.97 | 0.88 | 0.86 |
| This study | Mask R-CNN | 0.96 | 0.80 | 0.88 | 0.70 |

*Note.* Our approach exhibits comparable performance. Note that Earth-based observation has several advantages over the Cassini data, including data volume and geosynchronous observation.

rapidly and demonstrates a high validation accuracy of 0.96 with a precision of 0.80. We note that our accuracy, recall, precision, and IoU scores are comparable to similar cloud identification and segmentation works on Earth. LeGoff et al. (2017) used a convolutional neural network for this purpose, achieving a precision score of 0.81 and a recall of 0.75. Similarly, L. Li et al. (2021) fused high-resolution satellite data from multiple sensors to achieve an average IoU of 0.9 and an accuracy of over 0.95. Francis et al. (2019) also demonstrated the efficacy of U-Net, another popular semantic segmentation architecture, recording a 0.91 accuracy. Lopez-Puigdollers et al. (2021) take advantage of pre-labeled data sets for Landsat-8 and Sentinel-2 to achieve an accuracy of 0.94 with a fully connected approach. Fabel et al. (2022) and Gonzales and Sakla (2019) use to pre-trained U-Nets with ResNet34 backbones, achieving IoU values of around 0.8. We summarize these other studies in Table 1. Our work on Titan demonstrates similarly high scores despite the difference in image quantity and quality compared to terrestrial remote sensing.

Although our study uses a subset of the full Cassini ISS data set, we do not anticipate that this introduces bias into the model. Our data set is large enough to contain a wide variety of clouds that sufficiently represents the broader Cassini set. We can once again liken this problem to terrestrial cloud tracking. Although Earth clouds are indeed highly variable, with no two the same, nevertheless training on a subset of clouds from 1 year across the Earth would be highly applicable to recognizing clouds in other years.

Numerous planetary science projects may also benefit from this technology. Image data sets for cloudy worlds including Jupiter, Uranus, and Neptune are all also available in NASA's Planetary Data System, and have seen few, if any, deep learning projects (Chanover et al., 2022). Luz et al. (2008) demonstrate an early application of automated cloud tracking on Jupiter, but their approach could be improved with modern deep learning algorithms. Caille et al. (2022) already demonstrate that k-means clustering is effective for clouds on Mars, noting that their approach might be further improved by deep learning. Mars has also seen several other applications of machine learning to rover missions (Bajracharya et al., 2008; Estlin et al., 2007; Mishkin et al., 1998) and in the recent Ingenuity helicopter (Tzanetos, Aung, et al., 2022; Tzanetos, Bapst, et al., 2022; Verma et al., 2023), emphasizing an increased interest in onboard models. These existing applications might benefit from models specialized to specific scientific applications such as ours. Similarly, Machado et al. (2022) highlight their intention to apply deep computer vision to their cloud tracking efforts on Venus. The effectiveness of our approach, with limited data and computational resources, should give encouragement to future researchers who face similar limitations that fundamental and important scientific information can be readily extracted. Our approach demonstrates that state of the art deep learning models are feasible tools even for researchers who may not have access to significant computational resources.

In contrast to manual approaches, machine learning techniques have the potential to enable precise tracking of clouds across many years of images, processing large image quantities significantly faster than any human. Thus, such techniques can save planetary scientists countless hours by automating scientific tasks such as cloud tabulation and mensuration. They also stand to further the state of the art by enabling the recognition of subtle trends and patterns that have never before been discernible, but emerge only after systematic quantitative analysis. Provided fast masking algorithms allow researchers to focus their analyses on relevant metrics such as cloud areas, centroids, aspect ratios, and other useful metrics. Having detailed cloud shape outlines may permit comparison with regional, mesoscale cloud models such as TRAMS (Barth & Rafkin, 2007, 2010; Rafkin & Barth, 2015; Rafkin & Soto, 2020; Rafkin et al., 2022) We also expect that researchers will apply their own domain knowledge to labeling such data sets for use in training deep learning models, enabling their deployment for other specific problems.

In addition to processing data after missions are complete, these models may be beneficial for in-flight data reduction as well. One of the most significant problems facing robotic missions is downlink bandwidth: only limited quantities of information can be transmitted back to Earth, especially for more distant missions using the Deep Space Network (DSN). As spacecraft cameras become more advanced, with larger detector arrays and increased dynamic ranges in sampling, increasingly large image files must be compressed and transmitted. Without a means of identifying useful images, a mission is currently compelled to either store and eventually transmit all of them over a long time period (as with New Horizons), or else to transmit only a subset of the data. This bandwidth bottleneck currently affects missions to every world in the Solar System, including major future missions such as Europa Clipper (Phillips & Pappalardo, 2014) or Jupiter Icy Moons Explorer (JUICE) (Grasset et al., 2013). Similarly, the upcoming DAVINCI mission to Venus will already deploy a Generative Adversarial Network (GAN) for its Compact Ultraviolet Imaging Spectrometer (CUVIS) that will "generate a reduced data set and help flag and prioritize full-resolution data" (Garvin et al., 2022). This demonstrates how some missions might also use on-board image segmentation models to determine which images are valuable and which to discard before transmitting. Our work provides a step toward addressing these unique opportunities and challenges in various planetary science contexts, including imaging of Titan. Decreasing transmission volume and processing speed becomes increasingly relevant given the anticipated influx of Titan image data due to several key endeavors in the coming decade (e.g., Dragonfly (Barnes et al., 2021)). Although our model may not yet be suitable for implementation on a spacecraft (depending on available computing power), it demonstrates that this is likely to be a feasible approach in the future.

Besides data processing for close-up space missions, extracting cloud feature parameters from more distant telescopic observations may also be possible. The recently launched James Webb Space Telescope (JWST) possesses an angular resolution as high as 0.04 (with NIRCam), which is capable of capturing images of Titan at modest spatial resolution (~220 km/pixel) (Nixon et al., 2016). Several ground-based telescope projects are also in development that will offer higher resolutions, including the European Space Agency's Extremely Large Telescope (EELT), which is anticipated to be commissioned in ~2027, the 25 m GMT (Giant Magellan Telescope) in Chile, and possibly also the 30 Meter Telescope (TMT) which is currently searching for a site. With 25–39 m apertures and adaptive optics, these observatories will possess a resolving power capable of capturing medium resolution images of Titan, up to 200 × 200 pixels. Furthermore, these systems will provide an improved spectral range resolution that is critical for cloud characterization. This will significantly increase the availability of resolved cloud data of Titan and other worlds.

Additionally, our work is intended to promote a reevaluation of existing expectations of the difficulty and time commitment to utilize machine learning tools in planetary science research. Though training state-of-the-art architectures from scratch may exceed the data, temporal, and computational resources of most planetary science projects, our transfer learning approach demonstrates that top of the line models are within reach. While our work addresses only the sole problem of cloud identification, there are many other image analysis problems that may be suitable for application of transfer learning methods, as well as more sophisticated image processing techniques. In particular we suggest that wider application to surface features identification as well as atmospheric discrimination may prove to be not only tractable, but also yield novel results.

## 5. Conclusion

In this paper we presented a novel approach to Titan cloud identification using transfer learning to effectively label the individual pixels of images of Titan. Despite Titan-specific challenges such as data availability and quality, our results are comparable to similar studies on Earth. We extracted key metrics including cloud areas and centroids that may be prohibitively time consuming to compute manually, and analyzed the efficiency of our technique. We demonstrated how our approach can elucidate trends in cloud development over the course of the Cassini mission.

Given upcoming missions to Titan and advances in ground-based telescopes, there is a need for robust, proven models that can process high-quality and large-quantity data. Even now, 7 years after the mission concluded, there is much to be learned from novel processing of Cassini data at scale. We show that deep learning techniques can be significantly more efficient than manual tabulation, especially given increasingly large data sets as spacecraft technology improves. We hope that this work motivates other planetary science researchers to implement deep models with transfer learning for other problems in planetary science, so that the field as a whole might enjoy the benefits of transfer learning for computer vision.

## Data Availability Statement

All code is available on Zenodo (https://zenodo.org/records/14009768) (Yahn et al., 2024a) and on Github (https://github.com/zacharyyahn/TitanCloudsSegmentation) (Yahn et al., 2024b). Images, backplanes, models, and all other files are available for download on Zenodo (https://zenodo.org/records/13988492) (Yahn et al., 2024c).

## References

**Acknowledgments**
Thanks to NASA Langley Research Center for enabling this project by providing funded access to Google Cloud Platform, and to Charles Liles for his assistance with setting up the Google Cloud environment.